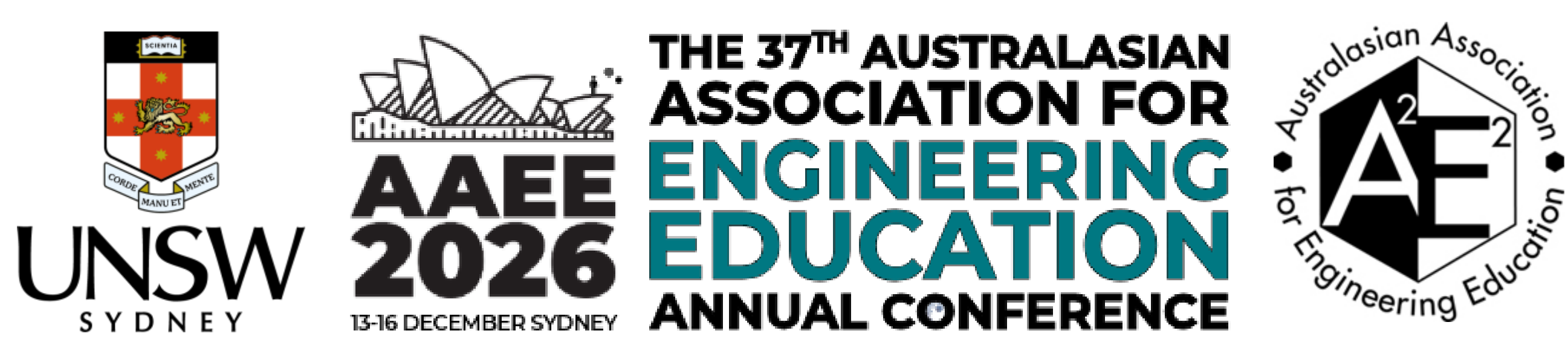

# Making the Invisible Visible: A Framework for Reflective AI use in Software Engineering Education

Ali Shakiba[a]; and Thomas Chaffey[a].
*The School of Electrical and Computer Engineering, The University of Sydney, Sydney, NSW, Australia, 2008[a]*
Corresponding Author Email: ali.shakiba@sydney.edu.au

## ABSTRACT

### CONTEXT

Generative AI (GenAI) is increasingly embedded in software engineering education, supporting activities such as requirements development, design exploration, documentation, and prototyping. However, educators often have visibility only into final artefacts, with limited insight into how students evaluate, verify, and refine AI-generated outputs during the learning process. This creates challenges for assessing evaluative judgement and responsible AI-assisted practice.

### PURPOSE OR GOAL

This paper introduces the AI Journal, a structured reflection framework designed to make student-GenAI interaction visible in first-year software engineering education. The framework aims to support process-aware learning by capturing not only what students produce, but also how they supervise, verify, and adapt AI-generated outputs throughout software engineering tasks.

### APPROACH OR METHODOLOGY/METHODS

The AI Journal was deployed in a first-semester software engineering course with approximately 140 students. The framework combines two complementary components: execution tracking, which records prompts, outputs, intent, and interaction context; and cognitive auditing, which captures verification strategies, intervention decisions, confidence judgements, critical learning moments, and reflections on AI-supported work. Journal completion was embedded within weekly learning activities as a hurdle requirement and used to support both student reflection and instructional insight.

### ACTUAL OR ANTICIPATED OUTCOMES

The AI Journal enabled visibility into aspects of student learning that were not observable through artefact-based assessments alone. AI Journals suggested variation in verification practices, intervention strategies, and perceptions of AI-supported work. Critical learning moments frequently occurred when students evaluated the contextual suitability, feasibility, and requirements alignment of AI-generated outputs rather than when identifying obvious errors. The collected data also enabled instructors to identify common misconceptions and informed subsequent teaching activities.

### CONCLUSIONS/RECOMMENDATIONS/SUMMARY

The AI Journal demonstrates a practical approach for making AI-assisted learning processes visible without restricting student use of generative AI tools. By foregrounding verification, intervention, and reflection, the framework shifts attention from product-focused assessment towards evaluative judgement and responsible AI-assisted practice. The approach is lightweight, model-agnostic, and adaptable to educational settings where understanding the process of AI-supported work is as important as evaluating the final artefact.

## Introduction

Generative AI (GenAI) tools are rapidly reshaping how students engage in software engineering education (Dwivedi et al., 2023; Kasneci et al., 2023; Siládi, 2025), supporting activities such as requirements development, design exploration, documentation, and code generation (Keuning et al., 2024; Prather et al., 2023). While these tools can improve productivity, they also introduce a significant pedagogical challenge: the invisibility of the students' decision-making during AI-assisted work. Educators are often left to evaluate only the final artefact, with limited insight into how students prompted, evaluated, and refined AI-generated outputs throughout a task (Boud & Falchikov, 2007, p. 3). This visibility gap is particularly important in first-year software engineering courses, where students are simultaneously developing both disciplinary knowledge and evaluative judgement (Bearman et al., 2024).

Several approaches have attempted to make learning processes more visible. Reflective journals and e-portfolios support metacognitive development and documentation of learning experiences (Beckers et al., 2016), but were not designed to capture AI-mediated decision-making as it occurs. Think-aloud protocols provide rich insight into learners' cognitive processes during problem solving (Prather et al., 2024), yet are difficult to implement at classroom scale. Emerging tools such as Promptly (Denny et al., 2024) and studies based on GitHub Copilot telemetry (Shihab et al., 2025) can record prompts, suggestions, and patterns of AI use, but depend on platform-level data access that is rarely available in educational settings and provide limited insight into the reasoning behind students' actions. Similarly, institutional disclosure approaches, including the AID Framework (Weaver, 2024) and post hoc AI-use declarations, strengthen attribution and transparency but focus primarily on reporting outcomes rather than documenting decision-making during AI use. Gonsalves (2025) also highlights the compliance challenges associated with self-disclosure, noting that students may selectively report AI use depending on context and perceived expectations. Collectively, these approaches can reveal what AI produced or whether it was used, but provide limited visibility into how students evaluated and adapted AI-generated outputs during the learning process. Shneiderman (2020) identifies such human oversight as central to responsible human-in-the-loop AI systems, yet this dimension remains poorly instrumented in educational contexts.

To address this gap, we developed the AI Journal, a structured reflection scaffold designed to document both the technical and cognitive dimensions of the student-AI interactions. The framework records prompts, outputs, and interaction context alongside students' verification strategies, confidence judgements, intervention decisions, and reflections on AI-assisted work. To support broad adoption, the design is intentionally model-agnostic and lightweight, requiring neither specialised infrastructure nor access to platform-level interaction logs.

This paper reports the deployment of the AI Journal in ELEC1005 (Introduction to Software Engineering), a first-semester, first-year course enrolling ~140 students at the University of Sydney. We describe the design of the framework, its integration into assessment, and preliminary insights derived from the resulting interaction data. Through this implementation, we explore how structured documentation can make AI-assisted learning process more visible and support the development of responsible engineering practice.

## Context: The ELEC1005 and Pedagogical Setting

The AI Journal was implemented in a first-semester, first-year software engineering subject designed for students with no assumed prior programming experience. The subject introduces foundational software engineering concepts (Kumar et al., 2024, p. 237), including requirements analysis, system design, prototyping, documentation, testing, and iterative development. Students primarily work with low-code and no-code technologies, such as Microsoft PowerApps, SharePoint Lists, and Figma, alongside modelling and documentation artefacts including UML diagrams. A key pedagogical objective of the subject is to expose students to the software development lifecycle (SDLC) before introducing substantial programming content. Consequently, assessment emphasised requirements interpretation, design reasoning, communication, and justification of engineering decisions rather than algorithmic implementation. This creates opportunities for GenAI-supported activities such as ideation, specification refinement, documentation, and design exploration. The cohort comprises approximately 140 students from diverse academic backgrounds, including engineering, science, and non-technical disciplines. While many students have prior informal experience with tools such as ChatGPT, few have received training in verification, critical evaluation, or responsible AI use.

Within this context, GenAI was treated as a legitimate tool that students could incorporate into learning and project activities. The educational challenge was therefore not whether students used AI, but how that use could be made visible in ways that supported reflection, feedback, and the development of evaluative judgement. The AI Journal was introduced as a structured mechanism to document and reflect on AI-assisted work throughout the semester.

## The AI Journal Framework

The AI Journal (available at https://ashakiba.com/ai-journal) is a model-agnostic scaffold for documenting student-GenAI interactions during software engineering tasks. It was designed to capture not only the observable interactions with AI systems but also the reasoning students apply when evaluating and using AI-generated outputs. The design encourages explicit judgement and reflection by asking students to self-report their acts of verification, intervention and justification.

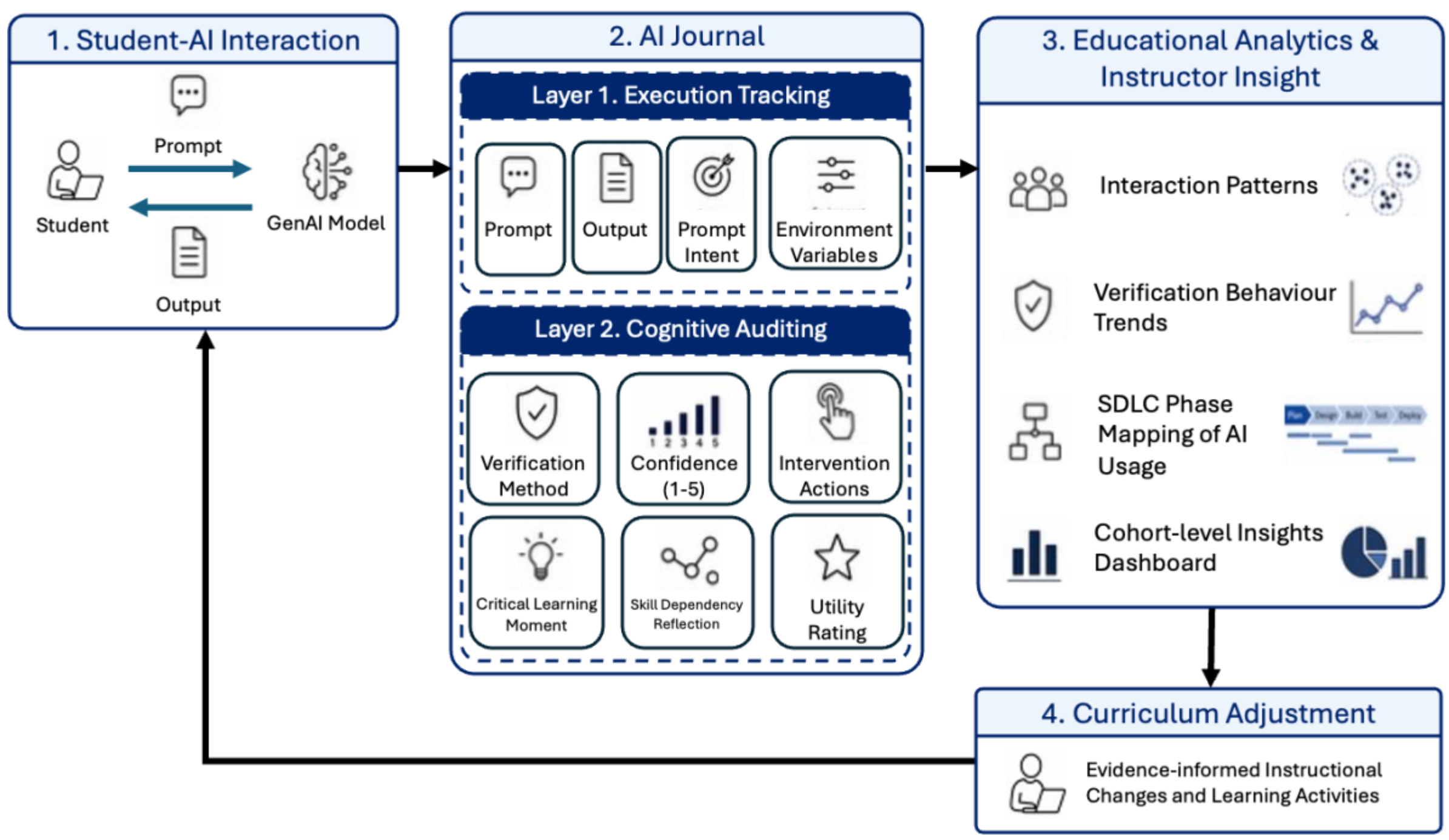

**Figure 1 – AI Journal as a human-in-the-loop system linking student-AI interaction, reflective capture, and instructor analytics.**

The framework consists of two complementary components: (i) Execution tracking and (ii) Cognitive auditing (Moon, 2013, Chapter 2; Shneiderman, 2020). Together, they provide a record of both the interaction itself and the student's evaluation of it. Figure 1 presents the AI Journal as a human-in-the-loop learning architecture linking student activity, reflective capture, and instructional analytics. Execution tracking records the interaction between the student and the GenAI system, providing sufficient context to understand how an AI-generated artefact was produced and how the interaction evolved over time. Students record: (1) **Prompt**: the student-authored prompt provided to the GenAI system, (2) **Output**: the generated response returned by GenAI, (3) **Prompt intent**: the purpose of the interaction, such as concept clarification, design exploration, specification refinement, or debugging, and (4) **Environment variables**: contextual information, including the AI platform, model version, and interaction environment.

These fields establish a traceable record while remaining independent of any particular AI platform or provider. Cognitive auditing captures how students evaluate and respond to AI-generated outputs. Students record:

- **Verification method**: how the output was evaluated, such as logical reasoning, prior knowledge, consultation of external sources, testing, or manual inspection.
- **Confidence level (1-5):** self-reported confidence in evaluating the correctness of the output.
- **AI intervention for correction**: actions taken when limitations or errors were identified, including prompt refinement, regeneration, additional constraints, or manual modification.
- **Critical learning moment**: a description of situations where the student challenged, rejected, or overrode an AI-generated response and the knowledge used to do so.
- **Skill dependency check**: reflection on aspects of the task that would have been difficult to complete without AI assistance.
- **Process evolution:** reflection on how the student's understanding changed during the task.
- **Overall utility score (1-5):** perceived usefulness of the interaction for task completion.

These fields focus on activities that are typically absent from artefact-based assessment and AI-use declarations, yet are central to the effective use of AI-assisted workflows in education. The framework relies on student-authored reflection, making verification, judgment, and intervention explicit elements of software engineering practice rather than by-products of AI use.

## Implementation and Practical Realities

The AI Journal was deployed in ELEC1005 (Introduction to Software Engineering) across a cohort of approximately 140 first-year students over a single semester at the University of Sydney. It was embedded directly into weekly design and development activities, ensuring that reflective documentation occurred as part of normal coursework rather than as a separate exercise.

### Onboarding and initial alignment

Before formal use, students completed a structured onboarding sequence designed to establish shared expectations around responsible AI use and reflective practice, including three components. First, students participated in scenario-based discussions using short comics developed by the teaching team (Available at https://ashakiba.com/projects/beyond-declaration/). These scenarios explored authorship, responsibility, over-reliance, and design accountability in AI-supported work. Students responded through Padlet and discussed differing interpretations of appropriate AI use. Second, students completed an academic integrity activity using structured polling questions covering permissible AI use, attribution, and data handling. This activity clarified institutional expectations and exposed common misconceptions about GenAI in assessment contexts. Third, students completed a guided practice exercise using the AI Journal during laboratory sessions. Students documented a small AI-assisted project, allowing them to become familiar with the framework and expected level of detail before its integration into the coursework.

### Integration into Weekly Workflow

The AI Journal was integrated into weekly coursework activities rather than treated as a standalone assessment item. Students documented AI interactions while undertaking tasks such as requirements analysis, prototyping, and design refinement, ensuring that reflection remained grounded in authentic engineering activities. To reduce friction, the collection of execution-level data (prompts and outputs) was partially automated where possible, while cognitive and evaluative fields such as verification strategies, confidence ratings, and intervention actions) remained student-authored. This preserved the pedagogical objective of encouraging reflection during, rather than after, AI use. Students were supported through university-approved AI platforms, including Microsoft Copilot and Cogniti.ai, but could use any approved tool. Where available, chat exports or shareable links were used to pre-populate interaction records to reduce manual transcription overhead.

Importantly, the AI Journal was implemented as a hurdle requirement rather than a graded assessment component. Students declared whether AI had been used in weekly activities, with journal submission required only when AI use was reported. This approach encouraged disclosure while reducing incentives to optimise responses for marks. Journal entries could also support follow-up discussions or oral questioning when questions arose about project contributions, team dynamics, or inconsistencies between reported processes and submitted artefacts. Periodic presentations of aggregated, anonymised journal findings were also shared with the cohort to demonstrate how reflections informed teaching and discussions of AI-supported practice.

### Staff Workload and Assessment Considerations

From an instructional perspective, reviewing AI Journal entries introduced additional workload due to the volume and variability. Entries differed significantly in depth, structure, and specificity, particularly during the early weeks of the semester. To reduce the workload, an internal toolset was developed to aggregate and explore journal data. These tools enable lecturers to identify common interaction patterns, verification strategies, and recurring points of confusion across the cohort. The purpose is not to automate assessment, but to support sense-making across large volumes of semi-structured reflections. Although still under development, early experience suggests they substantially reduce manual effort while enabling cohort-level insights.

## Preliminary Insights

Preliminary insights were derived through descriptive aggregation of structured journal fields and iterative review of qualitative entries by the teaching team during and after the 10-week implementation period. Qualitative entries were reviewed by the authors, and the interventions were derived from the field of critical learning moments. AI Journal entries provided visibility into student-AI interaction, including prompt framing, output verification, and student intervention strategies. Aggregated observations were periodically shared with the students to inform discussion and to guide instructional emphasis.

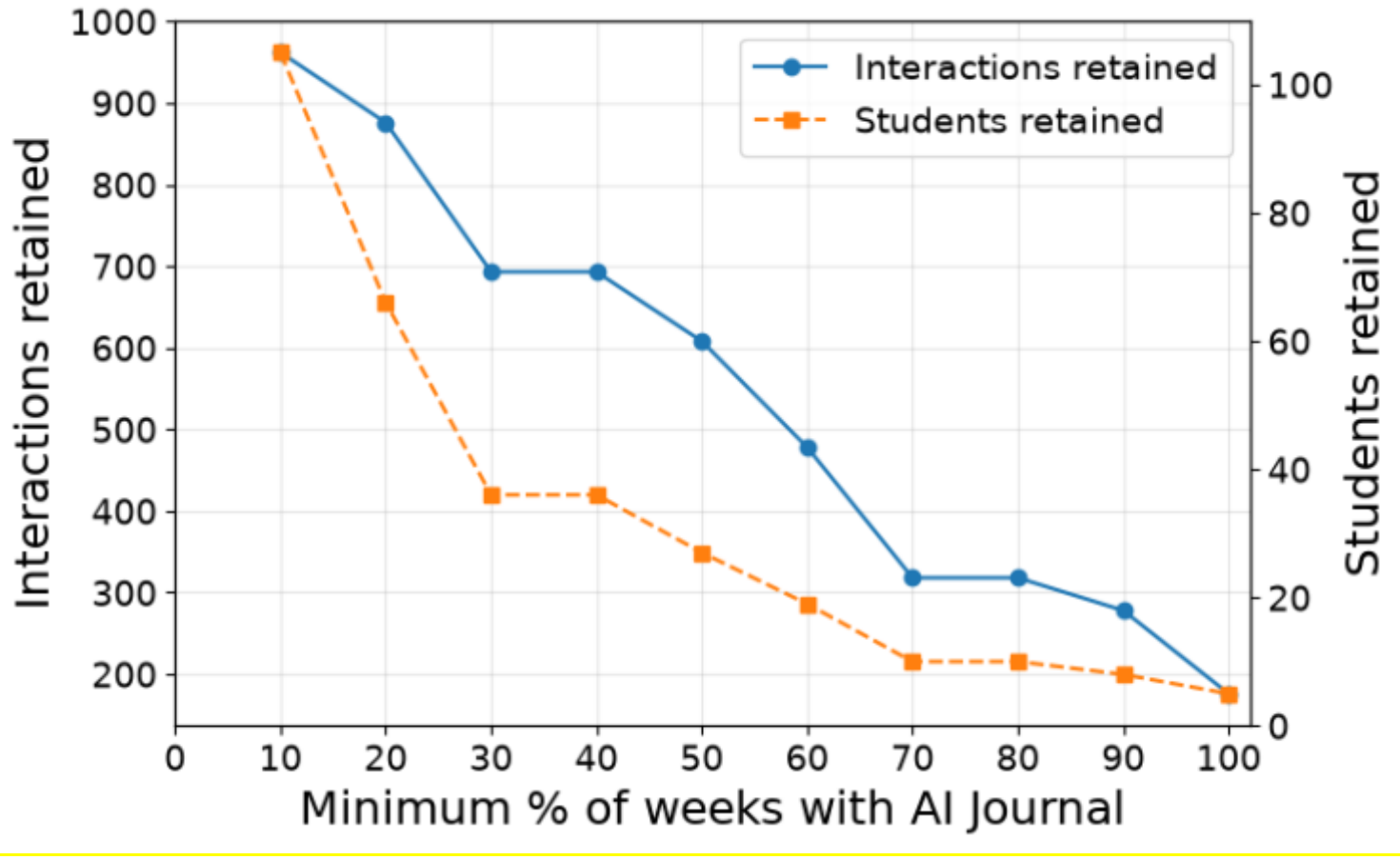

**Figure 2 – Retained interactions and participating students as a function of the minimum engagement threshold (percentage of active weeks).**

Data were collected over a 10-week period. The engagement threshold is defined as the minimum proportion of weeks in which a student either submitted an AI journal or explicitly declared no AI use. Figure 2 illustrates the relationship between this threshold and the number of retained students (right axis) and interactions (left axis). As expected, higher thresholds resulted in lower participation and fewer retained interactions. To improve analytical reliability, a minimum engagement threshold of 50% (≥5 of 10 weeks) was applied. This produced a final analytical sample of **27 students**, contributing **190 journals** and **608 interactions**. These students represented 25% of those who submitted at least one AI journal, while an additional **37 students** consistently reported no AI use. This relatively smaller sample reflects the hurdle design, which prioritised voluntary disclosure over maximising data yield rather than mandating participation. The retained dataset captures 63% of all logged interactions and 56% of total submitted journals (190 out of 336). The retained students reported high perceived utility of AI assistance (mean = 4.42/5) and high confidence in evaluating generated outputs (mean = 4.52/5). Students generally rated AI-generated outputs as high-quality, most often classifying them as “perfect” or requiring only minor modification. These ratings reflect students’ perception within the task context rather than an objective measure of correctness. The findings suggest differences in outcomes were driven less by perceived output quality and more by how students interpreted and validated generated content. Students were guided to document substantive generative interactions rather than incidental AI use. Informal feedback collected during laboratory sessions suggested that documenting a five-prompt interaction typically required less than 15 minutes, indicating a relatively low reporting burden.

## Verification Behaviour and Calibration of Confidence

Journal entries suggested an association between self-reported confidence (Dunlosky & Thiede, 2013) and recorded verification behaviour. Interactions involving verification were associated with higher mean confidence ratings than those without verification. This pattern suggests that the verification function not only serves as a corrective activity but also serves as an indicator of active engagement with AI-generated outputs. Rather than reflecting overconfidence, the relationship is consistent with the development of evaluative competence through experience.

## Critical Learning Moments

The "critical learning moment" field provided insight into the situations where students challenged, adapted, or rejected AI-generated outputs. These instances rarely involved obvious hallucinations or factual errors. Instead, interventions typically arose when outputs were plausible but misaligned with contextual requirements. Three recurring categories of intervention emerged: (1) **Contextual mismatch**: outputs that were too generic or insufficiently tailored to project-specific requirements or stakeholder needs, (2) **Feasibility constraints**: suggestions that exceeded platform limitations, implementation scope, or project constraints, and (3) **Requirements interpretation**: cases where generated solutions conflicted with assignment or system requirements.

These observations suggest that the primary form of evaluative judgement (Moon, 2013) exercised by students was not error detection in a narrow sense, but contextual adaptation and feasibility assessment, supporting the view that AI-assisted learning surfaces forms of engineering judgement that are typically invisible in final artefacts. Two representative examples illustrate these patterns. The first reflects requirements interpretation:

> *"The most important moment was when the AI proposed a solution that would have violated the assignment's core requirement to build a responsive, accessible mobile app... Instead of following the suggestion, I rebuilt it using Power Apps' responsive containers and relative sizing."*

This example is notable because the student recognised that a functionally valid suggestion was incompatible with an explicit project constraint and independently chose an alternative approach. The second example highlights feasibility constraints and emerging dependency on AI:

> *"Even though AI kept making invalid assumptions, it was still much faster to prompt then debug then prompt again than manually creating the formulas as I am not familiar with PowerApps syntax."*

Here, the student acknowledges both the productivity benefits of AI and a knowledge gap that reliance on AI may be reinforcing. Unlike the previous example, the student accepts the AI-mediated workflow despite recognising its limitations.

Together, these examples illustrate a spectrum of AI use: at one end, students subordinate AI outputs to their own judgment; at the other, convenience begins to outweigh independent expertise. Both patterns are pedagogically significant yet would be largely invisible in the final artefact alone.

## Perceived Skill Dependency and Role of AI

Student reflections consistently framed GenAI as a support tool rather than a substitute for task completion. AI was most commonly used for ideation, requirements structuring, terminology clarification, and exploration of alternative design approaches. At the same time, students repeatedly emphasised the need for human judgement when validating outputs, selecting solutions, and ensuring alignment with project constraints. This suggests an emerging distinction between generative support and evaluative responsibility. Rather than replacing learning, students often described AI as reducing the cognitive effort required for initial task structuring while preserving the need for domain expertise and validation.

## Adaptability and Educational Transferability

A key feature of the framework is its low technical overhead. It does not require access to platform-level logs, institutional monitoring systems, or specialised infrastructure. Instead, it relies primarily on student-authored records, supplemented where possible by automation capture of prompts and outputs. This reduces implementation barriers and supports deployment across a wide range of educational settings. The framework is also platform-agnostic. Students can use different GenAI systems while recording a common set of interaction and reflection fields, making the approach resilient to the rapid evolution of AI tools.

More broadly, the AI Journal is built around three transferable design principles: (1) **Visibility of interaction**: documenting prompts, outputs, and interaction context, (2) **Visibility of verification**: documenting how outputs were evaluated, tested, or validated, and (3) **Visibility of intervention**: documenting when and why outputs were modified, rejected, or refined. These principles transform otherwise hidden aspects of AI-assisted work into artefacts that can support reflection, discussion, feedback, and assessment. While the implementation described in this paper focused on software engineering education, the same principles may be applicable in disciplines where students use GenAI to create, evaluate, or refine artefacts, including engineering design, information systems, business analysis, technical communication, and other project-based learning environments where the quality of student reasoning is as important as the final product.

The primary contribution of the AI Journal is therefore not to regulate or restrict AI use, but to provide a practical mechanism for making evaluative activity visible. As institutions continue to integrate GenAI into teaching and assessment, approaches that foreground verification, intervention, and reflective judgement may provide a useful complement to artefact-based evaluation alone.

## Limitations

The AI Journal captures only interactions that students choose to disclose. As (Gonsalves, 2025) noted, students may selectively report AI use, meaning the framework provides visibility into reported rather than complete AI-supported activity. Non-participation is also difficult to interpret, as it may reflect limited AI use, disengagement, or deliberate avoidance of documentation. In addition, aggregate metrics may mask substantial differences in reflection quality, with superficially similar entries ranging from strong evaluative practice to indications of AI dependency. Strategies under exploration include exemplars, peer calibration activities, and AI-assisted feedback prompts. As AI tools become increasingly embedded in routine study activities, determining which interactions warrant documentation may also become more challenging. The AI Journal should therefore be viewed as a tool for reflection and pedagogical insight rather than a comprehensive audit of AI use. The framework was designed for contexts where AI use is permitted. In restricted-use settings, its role would shift toward a disclosure mechanism, reintroducing the compliance challenges noted above rather than resolving them.

## Conclusions and Next Steps

The increasing adoption of GenAI in higher education creates a challenge for assessment and learning design: educators can typically observe what students produce but have limited insight into how they evaluate, verify, and adapt AI-generated outputs. This paper introduced the AI Journal, a structured framework that captures both interaction activity and student evaluation of AI-supported work. Implemented in a first-year software engineering course, the framework combines execution tracking and cognitive auditing to document aspects of learning that are typically hidden from artefact-based assessment.

Our experience suggests that the value of the AI Journal extends beyond documenting AI use. Journal entries provided insight into how students verified outputs, responded to limitations, and assessed suggestions against contextual engineering constraints. The preliminary findings suggest that many important learning moments occurred not when AI produced obviously incorrect outputs, but when students were required to assess contextual suitability, feasibility, and alignment with project requirements. These forms of judgement are central to engineering practice yet are rarely observable through final artefacts alone.

From a practical perspective, the implementation demonstrates that process-oriented documentation can be integrated into existing coursework with relatively low technical overhead while preserving student access to GenAI tools. The framework is lightweight, model-agnostic, and adaptable, making it suitable for a wide range of educational contexts. The AI Journal is not intended as a compliance or surveillance mechanism. Because it relies on student self-disclosure, it provides insight only into reported interactions. Its value lies less in determining whether AI was used and more in understanding how students describe evaluating, challenging, and supervising AI-generated outputs. In this sense, the willingness to document and critique AI use may itself be indicative of the development of responsible professional practice.

Future work will examine AI Journal data across multiple cohorts and learning settings to better understand how verification practices, intervention strategies, and evaluative judgment evolve over time. We also plan to explore approaches for improving reflection quality, including exemplars, peer review, calibration activities, and AI-assisted feedback. Together, these directions aim to strengthen the role of process-oriented approaches in supporting responsible AI-assisted learning and professional development.

## Ethics statement

This study was approved by the Human Research Ethics Committee of the University of Sydney (Approval No. HREC 2025/HE001132 – substudy 022 v1a).

## Acknowledgements

This project was supported by an Education-Focused Academics SoTL Grant from the University of Sydney.

## Copyright statement